# Downward relativistic potential step and phenomenological account of Bohmian trajectories of the Klein paradox

M. Razavi[1], M. Mollai[1], S. Jami[1], and A. Ahanj[2, 3]

[1] *Department of Physics, Islamic Azad University, Mashhad Branch, Mashhad 91735-413, Iran*
[2] *Department of Physics, Khayyam Higher Education Institute (KHEI), Mashhad, Iran*
[3] *School of Physics, Institute for Research in Fundamental Science (IPM), Tehran, Iran*

Email: mollai.razavi@gmail.com

**Abstract**. The Dirac equation has been applied to fermions scattering from the downward potential step. The results show some particles do not fall off the edge of the step and reflect. Also, based on de Broglie-Bohm interpretation of quantum mechanics (Bohmian mechanics) and Bohmian trajectories we have resolved the problem. Lastly, a phenomenological study of the Bohmian trajectory of the Klein paradox has been discussed.

## 1. Introduction

Klein paradox (KP) is a well-known problem in the relativistic quantum mechanics that was first introduced by O. Klein [1], who calculated the Dirac equation for the problem of the finite square step. In the non-relativistic area, the phenomena of electron tunneling in the upward potential step with exponential damping are fully understood. Klein's results have shown that if the step height is in the order of electron rest mass, $eV \approx mc^2$, the step is almost transparent and more exotic sequence will appear as the step height is strong enough, so we will have a reflection coefficient larger than the incident beam and negative transform coefficient. Moreover, at the infinite limit of step height, the reflection coefficient becomes zero and transform is certain [2-4].

Some scientists explain this paradox using the interpretation of pair production and anticipate this for the potential well [2, 4, and 5]. But the Klein paradox which is discussed in terms of the single-particle Dirac equation in their work [2, 4, and 5] and the present paper, not in the contexts of the many-particle Dirac equation or Dirac quantum field theory so all the talk about pair creation is not relevant. This has already been done in the Greiner's book [6]. Greiner explains why the Bjorken and Drell account of the Klein paradox [4] is not entirely correct. E.g. the reflection coefficient is never larger than one. On the other hand, some researchers try to demonstrate this paradox in the context of the de Broglie-Bohm causal interpretation of quantum mechanics (Bohmian mechanics) and Bohmian trajectories [7, 8], and these works seem well-argued.

In the non-relativistic area, Bohmian mechanics presents a causal approach based on two elements: guidance formula, $\vec{v} = \vec{\nabla}S/m$ and the Newtonian equation that involves quantum potential (QP), $-\hbar^2 \nabla^2 R/2mR$, by considering wavefunction as $\psi = R exp(iS/\hbar)$ [9]. The guidance formula defines the dynamics of the physical system and QP in a Newtonian equation (extracted from the Schrödinger wave equation) tells us why it evolves in this way.

In relativistic area there is an expression for QP based on the Klein-Gordon equation, but it contains a couple of problems such as negative $J_0$ and space-like current. Hence, it does not contain a consistent picture [9, 10]. Consequently two schools have been created from the de Broglie-Bohm interpretation of quantum mechanics (see [11] for an excellent introduction to the subject], in which one has not considered the QP in their interpretation [12], while others have tried to generalize this concept and continue to find proper results from it [11, 13, 14]. Presently, in this area the mechanism which is introduced by Bohm [15] and has been successfully carried through in detail by Takabayasi [16] comprises consistent single-particle physical interpretation of Dirac's theory [7]. Holland applied this interpretation to the problem of KP and presented wholly reliable results that do not confound our intuition concerning this process [7].

The main goal of our paper is to consider the problem of scattering from the relativistic downward potential step may lead to a better understanding of KP phenomenon. In section 2 we have calculated the scattering coefficient of the downward potential step and have presented the analysis of asymptotic cases. In section 3 we have considered the downward step based on the view of Bohmian mechanics and calculated the Bohmian trajectories containing correlated results with the Holland interpretation of KP [7]. Section 4 is devoted to the discussion and we have argued the phenomenological account of Bohmian trajectories of relativistic steps and the KP case in the black hole and anticipating another form of Hawking radiation.

## 2. Relativistic downward potential step

Solution of Dirac equation for free spin-$\frac{1}{2}$ particles gives the normalized four component wave function as below [6]:

$$\psi = [(E+m)/2E]^{1/2} \begin{pmatrix} 1 \\ 0 \\ \dfrac{k^3}{E+m} \\ \dfrac{k^1 + ik^2}{E+m} \end{pmatrix} e^{-ik_\mu x^\mu} \quad , \quad E = k^0 = (\vec{k}^2 + m^2)^{\frac{1}{2}} \qquad (1)$$

This is normalized so that $j^0 = \psi^\dagger \psi = 1$. It is an eigenstate of the energy-momentum operator $\hat{p}^\mu$ corresponding to $k^\mu$.

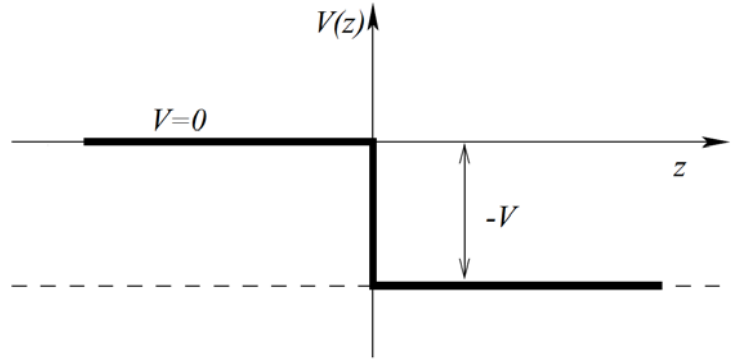

Figure 1. Downward potential step (attractive).

The incident, reflected and transmitted wave functions for the problem of downward potential step that is located on the $z$-direction (figure 1) is:

$$\psi_I = \begin{pmatrix} 1 \\ 0 \\ \dfrac{k}{E+m} \\ 0 \end{pmatrix} e^{-i(Et-kz)} \qquad (2)$$

$$\psi_R = A \begin{pmatrix} 1 \\ 0 \\ \frac{-k}{E+m} \\ 0 \end{pmatrix} e^{-i(Et+kz)} \qquad (3)$$

$$\psi_T = C \begin{pmatrix} 1 \\ 0 \\ \frac{q}{E+V+m} \\ 0 \end{pmatrix} e^{-i(Et-qz)} \qquad (4)$$

Where $k$ and $q$ are wave numbers as $k = (E^2 - m^2)^{1/2}$ and $q = [(E+V)^2 - m^2]^{1/2}$ for which $-V < 0$ and $q$ is always real.

As we expected, the only difference with the upward step is in the transmitted wave in which potential changes from $+V$ (repulsive) to $-V$ (attractive). The requirement of continuity at $z = 0$ implies that:

$$A = \frac{1-\gamma}{1+\gamma}$$

$$C = \frac{2}{1+\gamma} \qquad (5)$$

And $\gamma = \left( \frac{(E+V-m)(E+m)}{(E-m)(E+V+m)} \right)^{1/2}$.

The four-vector of the current can be calculated as $J^\mu = \bar{\psi}\gamma^\mu \psi$ in which $\gamma^\mu \equiv (\beta, \beta\vec{\alpha})$ are Dirac matrices and $\bar{\psi} = \psi^\dagger \gamma^0$.

$$\alpha = \begin{bmatrix} 0 & \vec{\sigma} \\ \vec{\sigma} & 0 \end{bmatrix} \quad , \quad \beta = \begin{bmatrix} I & 0 \\ 0 & -I \end{bmatrix}$$

$$\sigma_1 = \begin{bmatrix} 0 & 1 \\ 1 & 0 \end{bmatrix}, \quad \sigma_2 = \begin{bmatrix} 0 & -i \\ i & 0 \end{bmatrix}, \quad \sigma_3 = \begin{bmatrix} 1 & 0 \\ 0 & -1 \end{bmatrix}$$

So the currents associated with each of the elementary wavefunctions may be computed from $J = \psi^\dagger \alpha^3 \psi$ [6]:

$$J_I = \frac{2k}{E+m}$$

$$J_R = \frac{-2k}{E+m}|A|^2 \quad (6)$$

$$J_T = \frac{2q}{E+V+m}|C|^2$$

The proportion of reflected and transmitted flows can be obtained by dividing their currents into the incident current:

$$R = \frac{|J_R|}{|J_I|} = \left(\frac{1-\gamma}{1+\gamma}\right)^2$$

$$T = \frac{|J_T|}{|J_I|} = \underbrace{\frac{q}{k}\frac{E+m}{E+V+m}}_{\gamma}\left(\frac{2}{1+\gamma}\right)^2 = \frac{4\gamma}{(1+\gamma)^2}. \quad (7)$$

Here, against the upward step (the case of KP) $\gamma$ is positive-defined and can never be negative. Moreover, the reflected and transmitted coefficients are in the area of $0 \leq R, T \leq 1$ as the relation of $R+T=1$ is always required.

Analysis of asymptotic cases for particle energy and depth of potential gives interesting results:

In the limit of $V \to \infty$, $\gamma$ tends to $\left(\frac{E+m}{E-m}\right)^{1/2}$, which depends on the energy of the particle E, where there are two different cases. For $E \gg m$ as you could see $\gamma \to 1$ and considering the above relations $T \to 1$ and $R \to 0$. For low energies, $E \approx m$, $\gamma \to \infty$ as $T \to 0$ and $R \to 1$. As we expected, in the limit of $V \to 0$, coefficients behave conventionally so $T \to 1$ and $R \to 0$, meaning that for the shallow step all of the particles will pass. The general outline of results can be summarized as:

$$\begin{cases} 1)\ V \to \infty \Rightarrow \gamma \to \left(\dfrac{E+m}{E-m}\right)^{\frac{1}{2}} \quad for \quad \begin{cases} E >> m \Rightarrow \gamma \to 1 \Rightarrow \begin{cases} R \to 0 \\ T \to 1 \end{cases} \\ E \approx m \Rightarrow \gamma \to \infty \Rightarrow \begin{cases} R \to 1 \\ T \to 0 \end{cases} \end{cases} \\ 2)\ V \to 0 \Rightarrow \gamma \to 1 \Rightarrow \begin{cases} R \to 0 \\ T \to 1 \end{cases} \end{cases} \qquad (8)$$

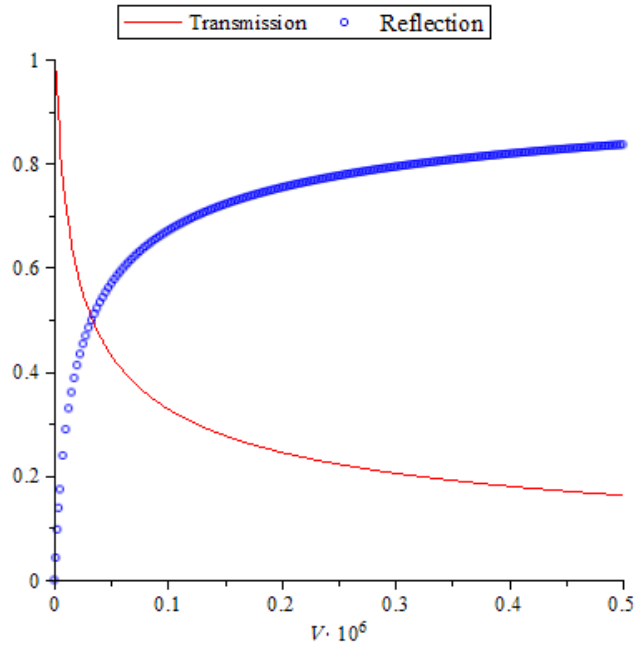

Figure 2. Scattering coefficient for low energy particle (m=1, E=1.001); with increasing depth of the attractive step, particles are less inclined to pass.

Since the step is *attractive*, the first case -in which we have complete transmission - seems logic and the third one also shows a conventional behavior. However, in the second case we could see almost total reflection from an attractive step, which is definitely non-classical. E.g. for the amount of $m=1$ and $E=1.01$; when $V$ is large enough the reflection coefficient is about 0.979 (see figure 2). This large reflection coefficient in this problem is not surprising and there are similar results in the non-relativistic area for the downward step also [17]. It has been shown that in contrary to classical mechanics, in which particles are reflected only from an upward step, quantum particles could be reflected from a downward step as well. Consequently, a quantum particle can be trapped for a long time (though not forever) in a region surrounded by downward potential steps, that is, on a plateau [17] (figure 2).

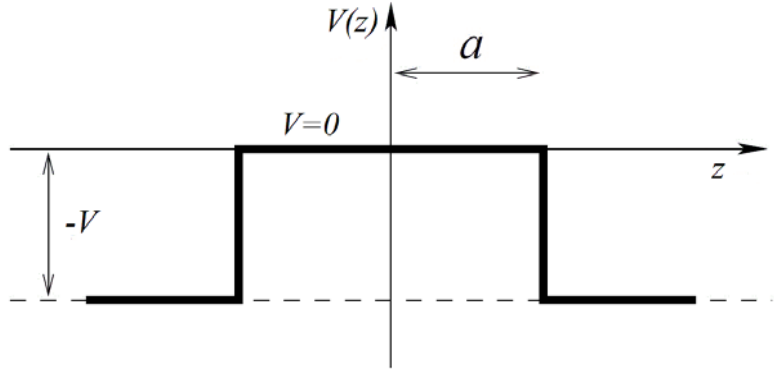

Figure3. Potential plateau.

These results have been affirmed in the presence of localized wave packets and for a soft step that goes to its maximum slowly [17]. The only additional condition in the later is the distance *L*, when the step hits its maximum point, it should be short enough in comparison with the electron Campton wavelength [17, 18]. As an example, the reflection coefficient for a shelvy step is zero when $L \to \infty$.

It is worth noting that there is an interesting difference on the reflection and transmission coefficients between relativistic and non-relativistic ones. In non-relativistic the reflection and transmission coefficients in scattering from a potential step with definite height both in the upward and downward case, are independent from the direction of the particle's path [17]. In other words, the coefficients are the same for the downward and upward step. The only difference occurs when the height of the step is higher than the energy of particles, such that we will have total reflection in the upward case. But, in the relativistic area this resemblance is no longer satisfied because in the upward step we have KP that involves challenging coefficients. However, in the downward case, except for non-zero reflection, we do not face any unusual result.

### 3. Holland approach and Bohmian trajectories

In section two, all calculations were based on the separate analysis of incident, reflected, and transmitted waves, whereas particles actually are under the effect of the superposition of $\psi_I$ and $\psi_R$ in $z < 0$. Based on the de Broglie-Bohm interpretation of quantum mechanics, taking the incident and reflected waves individually is meaningless because the actual wave function in this domain $(z < 0)$ is $\psi_{I+R}$ [7 and 9]. In the non-relativistic area, this perception is extensively discussed for potential step and barrier, in which consistent results have been obtained by using Bohmian trajectories and quantum potential (with both plane waves and localized wave packets) [19-22]. Holland compared the application of the de Broglie-Bohm theory of relativistic spin-1/2 particles to the Klein paradox [7]. Due to the conserved current implied by the Dirac equation, which is time-like and possesses a positive fourth component, a 3-velocity field could be presented as [7]:

$$v^i(x,t) = \frac{u^i}{u^0} = \frac{j^i}{j^0} = \frac{\psi^\dagger \alpha^i \psi}{\psi^\dagger \psi} \qquad (9)$$

For an arbitrary value of $V$, and writing $A = |A|e^{i\varphi}$, the probability density $j^0_{I+R} = \psi^\dagger_{I+R} \beta \psi_{I+R}$ and current $\vec{j}_{I+R} = \psi^\dagger_{I+R} \alpha^3 \psi_{I+R}$ in the region $z < 0$ are respectively [7]:

$$J^0_{I+R} = 2(E+m)^{-1}\left[(1+|A|^2)m + 2E|A|Cos(2kz-\varphi)\right] \qquad (10)$$

$$\vec{J}_{I+R} = \left(\frac{2k}{E+m}\right)(1-|A|^2) \qquad (11)$$

We intentionally use the constants and parameters similar to those presented in [7]. By substitution of the superposition current in (11), we could compute particles velocity $v_{I+R}$ in $z < 0$.

$$v_{I+R} = k(1-|A|^2)\left((1+|A|^2)E + 2m|A|Cos(2kz-\varphi)\right)^{-1} \qquad (12)$$

Similarly for $v_T$ In $z > 0$:

$$v_T = \frac{q}{E+V} \qquad (13)$$

By integration we were led to Bohmian trajectories:

$$(2kz-\varphi) + \varepsilon Sin(2kz-\varphi) = \omega t + c \quad , \quad z < 0 \qquad (14)$$

$$z = \left(\frac{q}{E+V}\right)t + c' \quad , \quad z > 0 \qquad (15)$$

Which $\varepsilon = \dfrac{2m|A|}{E+(1+|A|^2)}$ and $\omega = \dfrac{2k^2(1-|A|^2)}{E(1+|A|^2)}$

By choosing the constant as $c = c' = 0$, $m = k = \dfrac{1}{2}$ and $A = \sqrt{2} - 1$ (So $\varphi = 0$), so that it is equivalent to $V = 4\sqrt{2}$, we have the orbit:

$$\begin{cases} z + \dfrac{1}{2}\sin z = \dfrac{1}{2}t & (z<0) \\ z = \sqrt{\dfrac{17}{25}}\,t & (z>0) \end{cases} \qquad (16)$$

Also for velocities we will have:

$$v = \begin{cases} \dfrac{1}{2+\cos z} & (z<0) \\ \sqrt{\dfrac{17}{25}} & (z>0) \end{cases} \qquad (17)$$

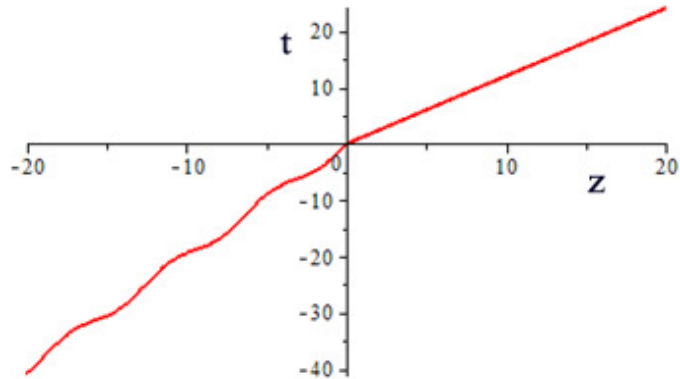

Figure4. Bohmian trajectories: All particles pass the step. Based on the causal view the reflected and transmitted coefficients in the plane wave representation do not accurately reflect the actual details of motion. The notion that the particle is 'incoming' or 'reflected' simply do not apply [9].

As you could see from the trajectories (see figure 3), track in $z<0$ does not have a permanent motion and there is an oscillation in the velocity, but in $z>0$, particles have a classical uniform straight-line motion. Considering this track, it has been concluded that in such phenomena, the expressions of reflection and transmission do not apply like classical mechanics, since particles are not reflected (figure 3) [7]. Now this wavefunction $\psi_I$ does not actually describe our private image of an initially free particle incident from $z = -\infty$, which interacts after only a certain period with step, for the reflected wave is already defined at $t = 0$ over the entire domain $-\infty < z < 0$ [7]. Similarly, the transmitted wave is defined for all $z > 0$ at $t = 0$. Based on the results it could be mentioned that having insufficient plane wave representation and having more real conditions for a free particle, we should use localized wave packets that give a more physical account [23].

## 4. Discussion: phenomenological account of Bohmian trajectories

As we mentioned in the introduction to invoke to the pair production interpretation to justify Klein paradox (KP) consequences [2, 4, and 5] is a misunderstanding since the problem is discussed in the context of single-particle Dirac equation and all the talk about pair creation is not relevant [6]. In [6] Greiner explains why the Bjorken and Drell account of the Klein paradox [4] is not entirely correct. Moreover when we apply Bohmian mechanics to the problem of relativistic upward step (KP) we do not encounter a paradox and we find that particles are coming out of the upward (repulsive) step.

There are some decisive indirect experimental evidences in graphene, which is a solid-state testing ground for quantum electrodynamics phenomena involving massless Dirac fermions [24-25]. In [24] it has been suggested that the transport characteristic through a p-n graphene junction can decide between the results obtained in the context of Bohmian mechanics [7-8] and the common Klein paradox theory, which uses pair production interpretation incorrectly.

We end our discussion by an annotation: There is something strange with the Bohmian trajectories of KP (figure 5). They emerge from the region with nonzero potential instead of moving into this region in the course of time [8]. Here we want to state that the results of the problem of fermions scattering from relativistic potentials (repulsive or attractive one) and their Bohmian trajectories may explain a new form of Hawking radiation. Recently, Phillips in [26] through others [27-29] expressed an assumption (based on the Dirac-Milne cosmological model [30]) that the universe has equal amounts of matter and antimatter that repel each other gravitationally, contrary to the prediction of general relativity [31]. It could solve simultaneously the puzzles of dark energy and dark matter through dramatic changes in the whole picture of physics. Although it should be experimentally tested, hence regarding this article, we could consider the event of the horizon in a black hole of matter (versus antimatter) as a relativistic step that is attractive (downward step) for matter and repulsive (upward step) for antimatter. Now we can interpret the underlying picture that Holland drew to explain KP in the de Broglie-Bohm theory of motion [7].

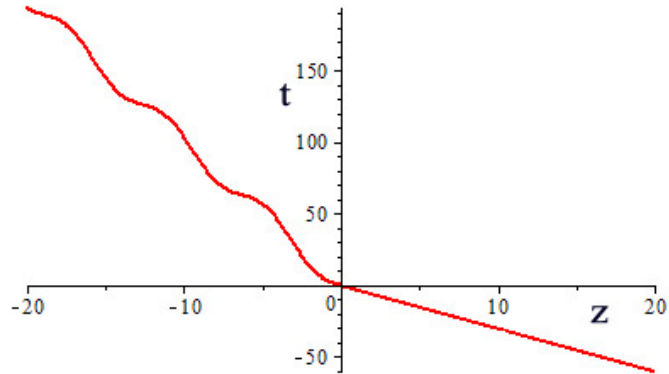

Figure 5. Bohmian trajectory of Klein paradox. The particles are coming out of the repulsive step.

As you can see in the figure 4, Holland finds that in the interior of the repulsive step, when $V > E + m$, we have negative flow which trajectories show and particles are coming out of the step. As a primary phenomenological interpretation, we could coincide this in black holes at the event of the horizon as follows: Due to vacuum fluctuations a pair of particle-antiparticle in the event of the horizon could be created [32] and through absorption of a particle by the black hole, the antiparticle will be ejected because of the repulsive force. So, the black hole will lose its energy that is stored in space-time. Since the black hole is repulsive for antiparticles, the exit of antiparticles from it coincides with the situation of KP, in which the Bohmian trajectory shows that particles are coming out of the step. So, this process leads to black hole evaporation and it is stronger than the Hawking radiation and could make the time of evaporating much smaller because in each pair production the antiparticle will certainly leave the black hole, and according to the amount of its energy, the black hole shrinks. Moreover, similar to Hawking radiation, in the appropriate condition the particle could escape. In this view, Bohmian trajectories are anticipating a phenomenon like the Hawking radiation in KP. Based on this description we are explaining a phenomenological interpretation of the Holland work and Bohmian trajectory on KP [7] as we suppose a black hole of matter for which is a downward step for antimatter (based on the Dirac-Milne cosmological model [30])and could have a flow of antiparticles exiting from it. An inverse picture does exist for an antimatter black hole. In another view, the above picture could support the pair production interpretation of KP somewhat; however, because of some opposing experimental evidence [24-25] it remains under debate. We recommend scientist do more rigorous mathematical calculations in curved spacetime for a justified answer in this area to falsify our results.

**Acknowledgments**

We would like to thank Professor M. Golshani, Dr. S. M. Saberi Fathi, and Dr M. Sarbishei for their useful advice and assistance.